\documentclass[12pt,superscriptaddress,preprint]{revtex4-2}
\usepackage[utf8]{inputenc}
\usepackage{amsmath}
\usepackage{graphicx}
\usepackage{float}

\newcommand{\Revised}[1]{#1}

\begin{document}

\title{Simulating COVID19 Transmission From Observed Movement: An Agent-Based Model of Classroom Dispersion}

%
%
%

\author{Yi Zhang}
\affiliation{University of Miami, Department of Physics, Coral Gables, FL}
\author{Yudong Tao}
\affiliation{University of Miami, Department of Electrical and Computer Engineering, Coral Gables, FL}
\author{Mei-Ling Shyu}
\affiliation{University of Miami, Department of Electrical and Computer Engineering, Coral Gables, FL}
\author{Lynn K. Perry}
\affiliation{University of Miami, Department of Psychology, Coral Gables, FL}
\author{Prem R. Warde}
\affiliation{Care Transformation, University of Miami Hospitals and Clinics, Miami, FL}
\affiliation{Data Analytics Research Team (DART) Research Group, University of Miami Hospitals and Clinics, Miami, FL}
\author{Daniel S. Messinger}
\affiliation{University of Miami, Department of Psychology, Coral Gables, FL}
\author{Chaoming Song}
\affiliation{University of Miami, Department of Physics, Coral Gables, FL}
\thanks{e-mail: c.song@miami.edu}

\begin{abstract}
Current models of COVID-19 transmission predict infection from reported or assumed interactions. Here we leverage high-resolution observations of interaction to simulate infectious processes. Ultra-Wide Radio Frequency Identification (RFID) systems were employed to track the real-time physical movements and directional orientation of children and their teachers in 4 preschool classes over a total of 34 observations. An agent-based transmission model combined observed interaction patterns (individual distance and orientation) with CDC-published risk guidelines to estimate the transmission impact of an infected patient zero attending class on the proportion of overall infections, the average transmission rate, and the time lag to the appearance of symptomatic individuals. These metrics highlighted the prophylactic role of decreased classroom density and teacher vaccinations. \Revised{Reduction of classroom density to half capacity was associated with an 18.2\% drop in overall infection proportion while teacher vaccination receipt was associated with a 25.3\% drop.} Simulation results of classroom transmission dynamics may inform public policy in the face of COVID-19 and similar infectious threats.
\end{abstract}

\flushbottom
\maketitle
%
%


\section*{Introduction}

\Revised{On March 11, 2020, the World Health Organization declared COVID-19 a pandemic  and called for coordinated mechanisms to support preparedness and response efforts across health sectors}\cite{zhu2020novel,wu2020nowcasting,dong2020interactive,world2021covid}. 
Predictive models can effectively inform policy to coordinate social policy responses to infectious pandemics\cite{giordano2020modelling}.  
The CDC has embraced the use of mitigation strategies in schools to allow communities to keep preschools and K-12 schools open\cite{cdc2019guidance,cdc2019operational}. \Revised{Here we focus on simulating infection responses to the novel coronavirus (SARS-CoV-2, which is responsible for COVID-19) based on observed interactions in a classroom setting}. To our knowledge, this is the first study to apply \Revised{Susceptible, Exposure, Infected, Recovered (SEIR)} to physically-defined interactions in these contexts. \Revised{Results may inform policy decisions related to vaccination, school attendance, and school closure and reopening. 
} 

\Revised{SEIR} models are used to quantify the spread of epidemics at a societal level\cite{he2020seir,yang2020modified}. 
SEIR models of COVID-19 quantify growth rate with respect to sample and population characteristics that may change with time\cite{warde2021linking}. 
Given their population focus, these models do not characterize individuals as transmission agents, and do not model their activity over time in a physical space inhabited by other agents. As SEIR models do not model granular transmission processes, they may be limited in the degree to which they speak to policy initiatives involving the repeated activity of individuals in physical space, such as schools. 

\Revised{While movement data collected by GPS~\cite{firth2020using, nouvellet2021reduction}, mobile phone records~\cite{gozzi2021estimating, muller2021predicting, chang2021mobility, aleta2020modelling}, and public transportation~\cite{kraemer2020effect} has been used to investigate the transmission of SARS-CoV-2 and other viruses, the spatial resolution of those sensors is  largely limited to ranges of meters to kilometers, which are unsuitable to the study of classroom and other enclosed space transmission.} Two recent reports demonstrate the potential of more granular analysis. One analysis modeled reported infection outcomes based on estimates of exposure duration and the physical density of individuals derived from published outbreak data\cite{tupper2020event}. This report contributed a model of infection saturation in small, relatively defined samples, but did not quantify transmission dynamics between individuals in physical space. Likewise, a recent model of transmission at a conference event quantified the importance of contact duration during which a specific pair of individuals were within an a priori (badge-based) physical range\cite{stehle2011simulation}. 
\Revised{Similar techniques have been used to study classroom transmissions~\cite{smieszek2019assessing}}. In contrast to the current effort, physical distance was not measured continuously nor was the relative physical orientation of individuals modeled. 

Data from 2020 indicate that young children may acquire COVID-19 (including confirmed, asymptomatic cases) in childcare settings and subsequently infect their household members\cite{lopez2020transmission}.  
Moreover, contact tracing revealed high levels of transmission between individuals of similar ages, including among children younger than four years of age\cite{laxminarayan2020epidemiology}. 
Thus there is evidence that young children may serve as vectors for COVID-19, Badge-based technology has been used to explore the temporal dynamics of transmission among high-school students\cite{smieszek2019assessing}, but it is not clear how physical motion facilitates infection dynamics in classrooms. The current study capitalizes on automated observations of pre-COVID-19 classroom physical interaction to model transmission between children and their teachers. It quantifies the dynamics of potential classroom outbreaks to help support efforts to mitigate outbreaks. To do so, we simulate the consequences of both half-capacity classrooms and teacher vaccinations on infection outcomes. These outcomes include infection saturation (percent and number of infected individuals) over time, time to the emergence of the first, second, and third symptomatic individual, and infection saturation levels when these symptomatic individuals emerge (a potential trigger of school closure).

\section*{Methods}

\subsection*{Data Sources and Calibration}
\begin{figure}[t]
\centering
\includegraphics[width=0.6\linewidth]{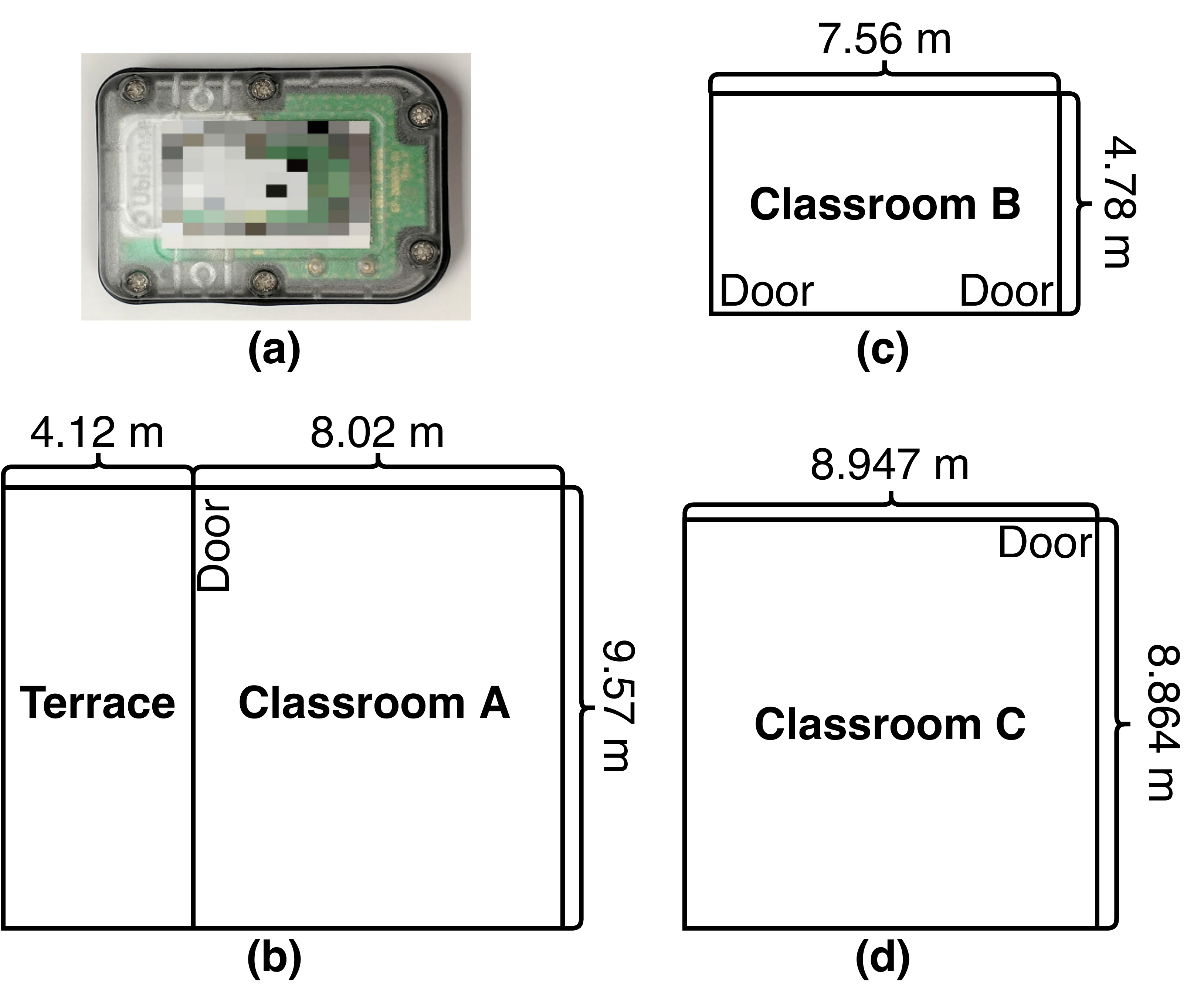}
\caption{(a) Ubisense tag worn by the teachers and children in the classrooms; Layout of Classrooms for (b) Class A; (c) Class B and (d) Classes C \& D. (Classes C \& D occupied the same classroom in successive years).}
\label{fig:layouts}
\end{figure}

Simulations were based on continuous real-time location tracking of individuals (students and teachers). A total of 34 observation periods occurred for four preschool classes housed in three classrooms in a large urban center in the United States. \Revised{Observations were conducted in inclusion classrooms for typically developing children and children with developmental disabilities contained in both a standard public county-funded school (Classes C and D) and a county-funded university preschool (Classes A and B). The range of represented class sizes (10-17) and teacher-child ratio are typical of preschool classrooms, as state regulations require preschool classrooms to have fewer than 20 children and teacher:child ratios of at least 10:1~\cite{bowne2017meta}.} Ultra-Wide Radio Frequency Identification (RFID) technology such as that embedded in child-worn Ubisense tags Fig.~\ref{fig:layouts}(a), allowed for efficient capture of individuals’ location and movement in the classroom indicating when children and teachers are in proximity. The Ubisense Dimension system was used to track individuals’ location at $2-4$ Hz to an accuracy of $21$ cm in the classrooms\cite{irvin2018automated}.  
The system consists of one sensor in each corner of the classroom, a dedicated server, and active tags worn by children and teachers. All children and teachers in the classrooms were tracked with two tags worn on their left and right hips (children wore a specially constructed vest and teachers wore fanny packs). The tags’ ultra-wide-band RFID signals were used to calculate the locations and orientations of both children and teachers in a three-dimensional space by means of triangulation and time differences in arrival. Specifically, each individuals’ locations were estimated as the centroids of their two tags while their orientations were calculated from the relative positions of the left and right tags.

Fifty 2-5-year-old children enrolled in four preschool classes and their 11 teachers participated in a study in which continuous RFID measurements of each individual’s location were collected. The classes were observed in academic years 2018-2019 and 2019-2020. \Revised{The last observation took place on 2/28/2020, prior to the impact of COVID-19.} Recordings were collected during observations of the entire school-day ($2.5-3.5$ hours), spaced 2-4 weeks apart in each class. Figure~\ref{fig:layouts}(b)--(d) plots the layout of the classrooms. Class characteristics are reported in Table 1.

\begin{table}[t]
\centering
\begin{tabular}{ |p{5cm}||p{2cm}|p{2cm}|p{2cm}|p{2cm}|  }
 \hline
 
 Class     & Class A &Class B&Class C1&Class D\\
 \hline
 Number of observations   & 13    &12 & 5 & 4\\
 Average data collection length&   3.1 hours  & 3.2 hours   &2.2 hours & 2.3 hours\\
 Total number of children (girls) &17 (8) & 10 (4)&  12 (5) & 11 (7)\\
 Total number of teachers    &3 & 2& 3 & 3\\
 \hline
\end{tabular}
\caption{\label{tab:stat}Sample characteristics by class.}
\end{table}

\subsection*{Ethics Declarations}
The study was conducted in accordance with APA ethical standards and was approved by the University of Miami Institutional Review Board (\#20160509). All teachers gave their informed consent and received \$100 for their participation. Parents gave their informed consent on behalf of their children and received \$75 for their participation. Teachers also received a classroom gift of their choice (e.g., a new book shelf). All research was performed in accordance with relevant guidelines and regulations.

\subsection*{Transmission Model}

\begin{figure}[t]
\centering
\includegraphics[width=0.8\linewidth]{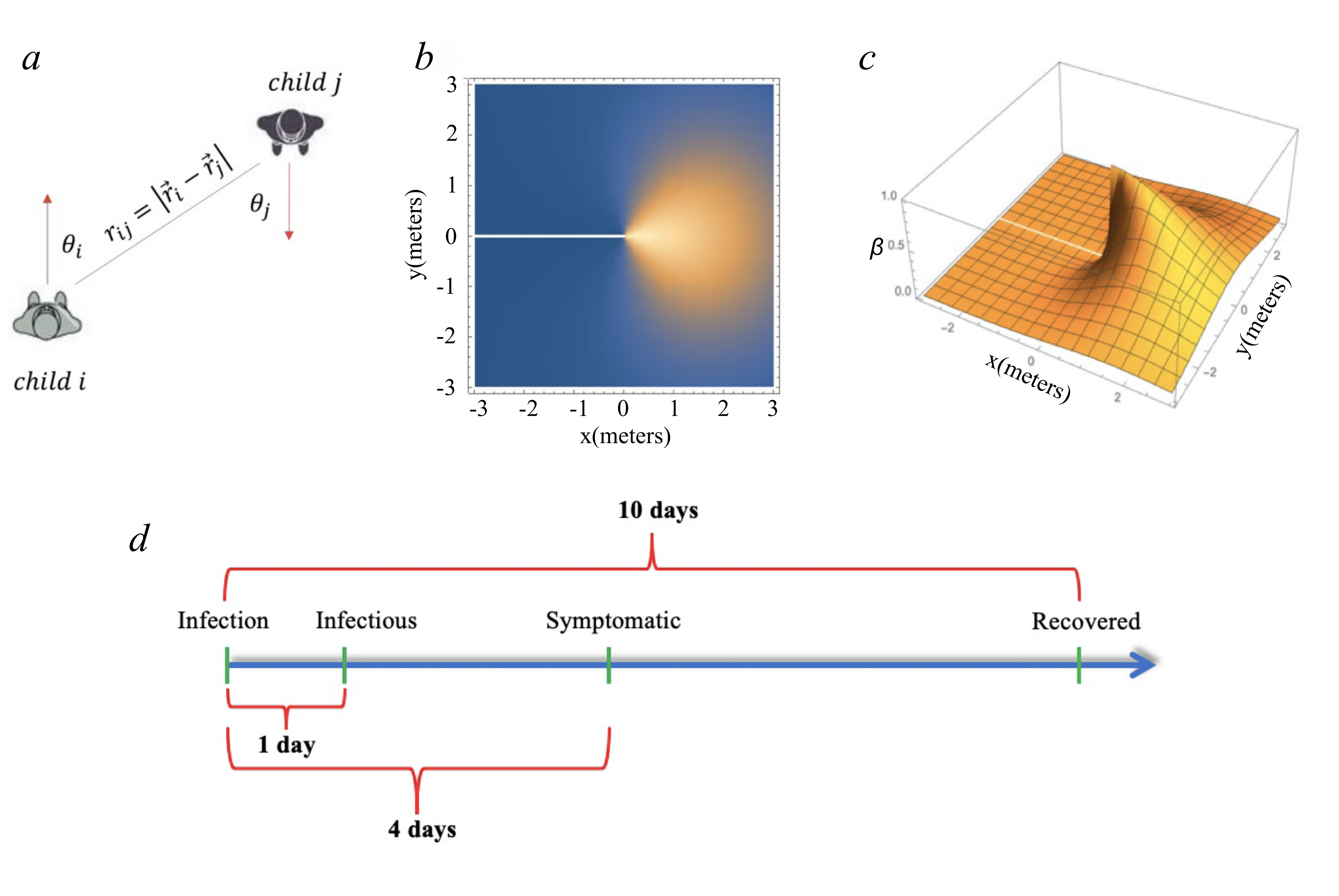}
\caption{(a) The definition of orientation for each pair of children, $r_{ij}$ is the distance between two children, and the orientation for a child $i$ is defined by angle $\theta_i$, \Revised{which is the angle between the direction being faced (red arrow) and the direct line between persons (blue line)}. A similar concept is applied to child $j$ to obtain $\theta_j$. (b) The heatmap and (c) the 3D plot of $\beta(r_{ij},\theta_i,\theta_j)$ with $x=r_{ij}\cos\theta_j$, $y=r_{ij}\sin\theta_j$ and $\theta_{i}=0$ where child $i$ is positioned at the origin and facing graph right. (d) The transition from infection to becoming infectious was $1$ day; the incubation period (from infection to symptomaticity) had a mean of $4$ days; the mean recovery time (from infection to infectious to recovered, i.e., non-infectious) had a mean of $10$ days.}
\label{fig:trans}
\end{figure}

We modelled the spread of \Revised{SARS-CoV-2} in the classrooms using a novel modification of the SEIR model. The typical SEIR model consists of susceptible, \Revised{exposed}, infectious, and recovered individuals where the susceptible individual becomes infectious after close contact with an infectious individual, with infection rate $\beta$, while infectious individuals recover at rate $\gamma$. For each individual i, the real-time classroom movement data records the position $(x_i, y_i)$ and the orientation $\theta_i$ for every second. To integrate the simplest SEIR model with the movement data, we develop a model to account for the dependence of the infection rate $\beta$ on distance $r_{ij}=\sqrt{(x_i-x_j)^2+(y_i-y_j)^2}$ \Revised{and the time difference $t=t_1-t_2$} between pairs of individuals ($t_1>t_2$) with interpersonal orientations $(\theta_i \& \theta_j)$, satisfying
\begin{equation}
   \beta(r_{ij},\theta_i,\theta_j,t) =\beta_{max}\exp\left(\frac{-r_{ij}^2}{2\sigma_{r}^2}-\frac{\theta_i^2+\theta_j^2}{2\sigma_{\theta}^2}\right)e^{-\lambda t},
\label{beta_def}\end{equation}
with $\sigma_r\approx 2m$, and $\sigma_\theta \approx 45^0$ where \Revised{$\beta_{max}$} is the maximum pairwise infection rate. That is, the rate of infection falls directly as an exponential function of growth in the square of the radius between the two individuals and their angular distance from face-to-face contact\cite{tang2008coughing}. A visual illustration of the distance $r_{ij}$ and orientations $\theta_i$ and $\theta_j$ is shown in Fig.~\ref{fig:trans}(a), and the heatmap and the 3D plot of $(r_{ij}, \theta_i, \theta_j)$ with $x=r_{ij} cos\theta_i$, $y=r_{ij} sin\theta_j$ and $\theta_i=0$ are presented in Fig.~\ref{fig:trans}(b) and (c), respectively. \Revised{A detailed discussion of the infection function is found in SM Section 4.C and 4.D. The $e^{-\lambda t}$ term accounts for temporal decay, modeling declining transmission over time $t$ which  characterizes the decay rate. For airborne transmission, $\lambda=0.34/\mathrm{hours}$~\cite{bazant2021guideline}, whereas droplet transmission decay is functionally instantaneous over a timescale of seconds (see SM Section 4.A for more details).}

To calibrate $\beta_{max}$ of the agent-based SEIR model, we calculate the average infection rate 
\begin{equation}
    \Bar{\beta}=\frac{N\iiiint \ \beta_{max} \exp \left( -\frac{x^2+y^2}{2\sigma_r^2} - \frac{\theta_i^2+\theta_j^2}{2\sigma_{\theta}^2} \right) \,dx\,dy\,d\theta_i\,d\theta_j}{A(2\pi)^2}
    =\beta_{max}\sigma_r^2\sigma_{\theta}^2\rho.
\label{eq_beta_bar}
\end{equation}
Note that the population infection rate $\rho_0$ is proportional to the density $\rho\equiv \frac{N}{A}$ whereas $\beta_{max}$ is an intrinsic parameter, independent of the social environment. Researchers have estimated the average daily infection rate at the meta-population level daily $\Bar{\beta}_{daily} \approx \beta_0$, where $\beta_0=R_0\gamma$ with $R_0$ and $\gamma$ being the reproduction number and the recovery rate , respectively.  Combining Eq.~(\ref{eq_beta_bar}), we calibrate our modeling parameter
\begin{equation}
    \beta_{max}=\frac{\Bar{\beta}_{daily}}{\sigma_r^2\sigma_{\theta}^2\rho_{daily}}.
\label{eq_beta_max}
\end{equation}
SARS-CoV-2 transmission has been modelled and analyzed in many literatures while we posit that the reported $R_0's$ variations in these works are largely due to virus evolution/mutation, as well as social, political, and environmental between each cohort\cite{bertozzi2020challenges,li2020early,kucharski2020early,abbott2020transmissibility,read2020novel,liu2020transmission}. We chose a conservative lower bound of  $R_0=2.0$ for our numeric simulations based on reported $R_0$ values, which typically fall within the $2.0-3.0$ range\cite{li2020early,ferretti2020quantifying,kucharski2020early,abbott2020transmissibility,read2020novel,liu2020transmission}. CDC guidelines have suggested defining close contact as being within $r=6$ feet of an infected person for a cumulative total of $T=15$ minutes or more over a 24-hour period\cite{cdc2019interim}. We thus estimate the daily population density $\rho_{daily}=\frac{N_c}{\pi r^2}\times(T/24\mathrm{hours})$ in Eq.~(\ref{eq_beta_max}) with average number of contacts $N_c \approx 10$\cite{mossong2008social}.

\subsection*{Numeric Simulations}

Based on the agent-based SEIR model, the spread of SARS-CoV-2 in classrooms can be simulated on a continuous-time basis. Here, infection only occurs during real time classroom observations that are separated by $24$ hours for each weekday (Monday through Friday) and $72$ hours between Friday and Monday. For each simulation, we start with a randomly chosen patient zero who is infectious at Day 0. For each time step $\Delta t=1$ second, we use the empirically measured location and orientation of each individual during the class time to compute the relative distance $r$ and orientations $\theta_i \& \theta_j$ for every pair of individuals $i$ and $j$ (including both teachers and children). Each susceptible individual is infected by infectious peers with a probability $\beta \Delta t$ calculated from all infected neighbours, where $\beta(r_{ij},\theta_{ij})$ is calculated based on Eq.~(\ref{beta_def}). The status of all individuals is updated every second based on the real-time location and orientation data. We illustrate the time dependent parameters related to infectious processes in Fig.~\ref{fig:trans} (d). Individuals become infectious $24$ hours after being infected\cite{cdc2019contact}. Once infected, individuals have a 75\% probability of becoming symptomatic\cite{johansson2021sars}; the transition to symptomaticity followed a poisson process with a mean of four days\cite{lauer2020incubation,guan2019china,li2020early}. Infected individuals become non-infectious with a probability $\gamma \Delta t$ where $\gamma=1/10$ ${days}^{-1}$ at each time step, reflecting a mean $10$-day duration\cite{cdc2019options}. We consider non-infectious individuals to be recovered, and assume no repeated infections. Each simulation continued until all \Revised{infected} individuals \Revised{are} recovered. All time periods include time in and out of the classroom. 

We applied the agent-based SEIR model to real-time location data from four preschool classes A-D. For each observation, we began with a single patient zero who was infected at $t = 0$ and simulated the observed classroom interaction over $60$ iterations. This process was repeated for each individual in the classroom (each was used as patient zero). For each simulation we output the (S, E, I, R) status of each individual and a mean individual hourly status is calculated over all simulations.

\subsection*{Public Health Scenarios}

We explored four public health scenarios pertaining to classroom density and teacher vaccination. To explore the role of classroom density, \Revised{SARS-CoV-2} infection was simulated for both full and half sized classroom scenarios. In the full class scenario, every student and teacher was simulated. In the half-class scenario, half the students in the class (rounding up for odd numbers of students) and one teacher were simulated. For example, if an empirical observation of Class A contained 18 children and 3 teachers, the half sized class simulations would involve 9 children and 1 teacher. The half-class simulations simultaneously model in which class sizes are administratively reduced and scenarios in which fewer students opt to attend class. 

At the time of writing, vaccination against COVID-19 was available for individuals above the age of 12, but not for younger individuals\cite{cdc2019interclinical}. Thus we conducted a parallel set of simulations assuming that classroom teachers were vaccinated. When vaccinated, we assumed a $85.8\%$ probability of vaccine effectiveness ($85.8\%$ probability of no infectivity) for each simulation run. The probability of effectiveness was based on the average efficacy of FDA-approved vaccines at the time of writing (Pfizer $97\%, $Moderna $94.1\%$, Johnson \& Johnson $66.3\%$)\cite{pfizer2021real,cdc2019moderna,cdc2019johnson}. For each of the four scenarios formed by the crosstabulation of classroom density and teacher vaccination levels, numeric simulations for each patient zero were repeated $60$ times. \Revised{There were no infections in simulations involved a vaccinated teacher (whose vaccination was effective) as the patient zero. For both full and half size class scenarios, the simulations are conditional on one infected person attending class. However, if we consider a school with all half-sized classes and double the number of classes, the chance of each class having a patient zero is reduced by a factor of two, which reduces the infections further.}

\subsection*{Policy-Focused Outcomes}

Numerical simulations were intended to inform decisions and policy designed to curb the impact of COVID-19 and other infectious agents in classrooms. We first calculated the time course of infection saturation in full-class and half-class scenarios with and without teacher vaccination over 28 days.  Next, we estimated the proportion of simulations yielding a  first, second, and third symptomatic individual, and the number of days required for their emergence. We reasoned that the emergence of a symptomatic child or teacher would be the first overt sign of COVID-19 infection in a classroom. Thus the number of days until a first symptomatic case emerged would index ‘damage done’ from a policy perspective. Likewise, the time to subsequent symptomatic cases would indicate the cost of ignoring a first infection. 

\begin{figure}[t]
\centering
\includegraphics[width=0.8\linewidth]{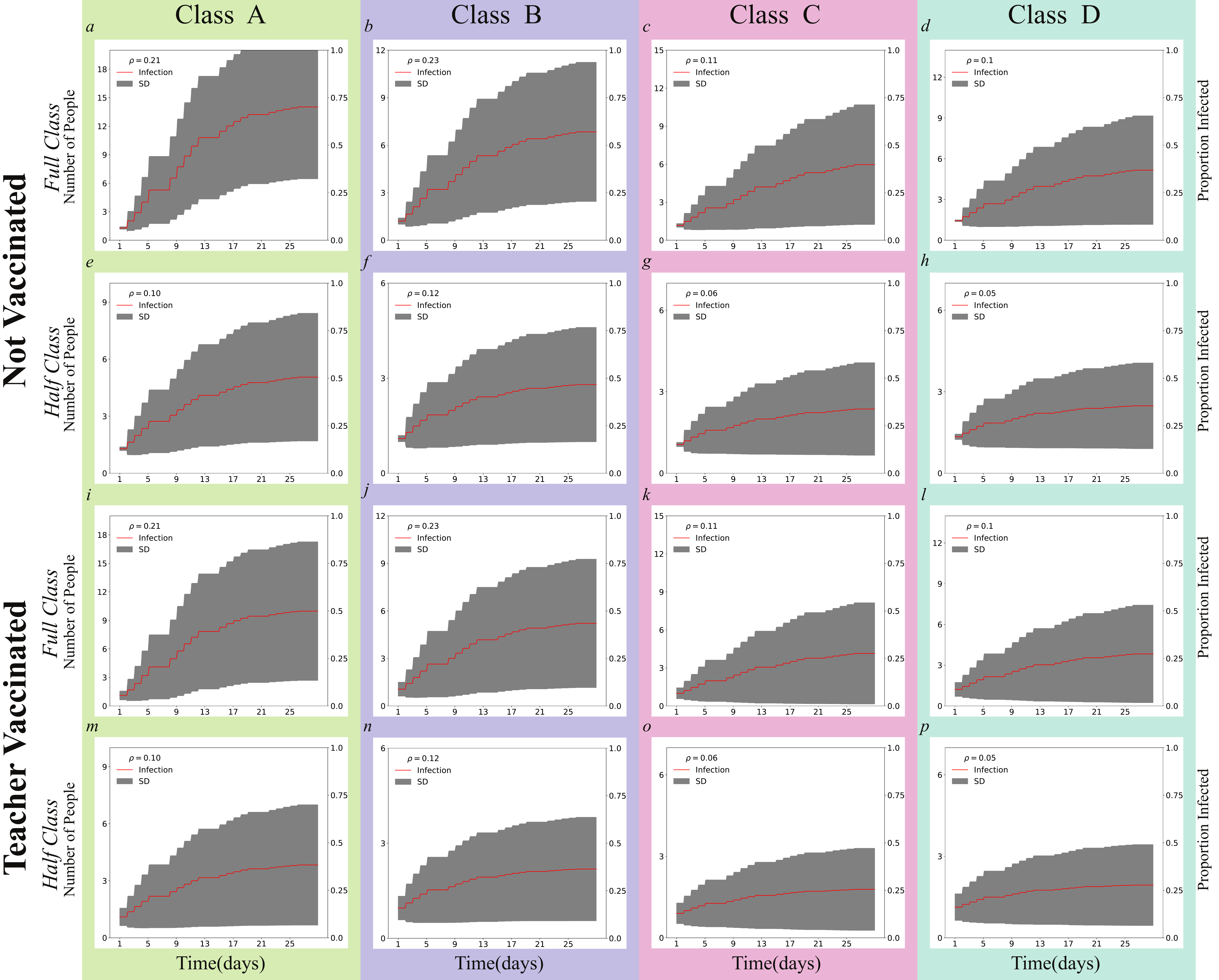}
\caption{\Revised{SARS-CoV-2} Infection over Time. The proportion (right) and total number (left) of infections over time in days . Trajectories are simulated for four classes for the full versus half-sized scenarios and not vaccinated versus teacher-vaccinated scenarios. The red curves and grey areas represent the mean and standard deviation of the proportion/total number of infections for all observations, respectively. $\rho$'s in each legend represent the classroom densities in the unit of people per $m^2$.}
\label{fig:infection}
\end{figure}

\section*{Results}

\subsection*{Infection Over Time}

To investigate the spread of \Revised{SARS-CoV-2}, we plot the proportion (and number) of infected individuals over one month in Fig.~\ref{fig:infection} for both full/half classes and non-vaccinated and teacher-vaccinated scenarios. The mean level (red line) and standard deviation of infection (grey area) over simulation runs is presented. The classroom population density, $\rho$, the number of individuals per square meter of classroom space is calculated for each observation (see SM Section 2 for more details). Figure~\ref{fig:infection} shows lower infection levels over time in half-sized than full size simulations. Likewise, the teacher vaccinated scenarios yield lower infection levels over time than the not vaccinated scenarios. In addition, Figure~\ref{fig:infection} suggests an association between the classroom density $\rho$ and infection levels over time. Specifically, scenarios with higher densities produced higher proportions of infected individuals. These findings suggest that classroom density plays an important role in controlling \Revised{SARS-CoV-2} spread. 


\subsection*{Transmission Likelihood and Saturation}
 
To quantitatively investigate the impact of classroom density, we characterize the infection patterns of the numerical simulations using the methodology of Tupper et al\cite{tupper2020event}. In the current model, the time-dependent infection number is determined directly by the agent-based simulation (see Fig.~\ref{fig:infection}), providing exposure duration and similar transmission measures\cite{tupper2020event}. Here we focus on the association between two transmission measures, saturation and average transmission rate. Saturation is the cumulative number of infected individuals (normalized by the total population) at the end of the simulation (final proportion infected). The transmission rate $\hat{\beta}$ is computed by averaging from Eq.~\ref{beta_def} at one second intervals across individuals. The product of the transmission rate $\hat{\beta}$ and the duration time $T$ reflects the likelihood of an individual being infected (see SM Section 2 for more details). 

\begin{figure}[t]
\centering
\includegraphics[width=0.8\linewidth]{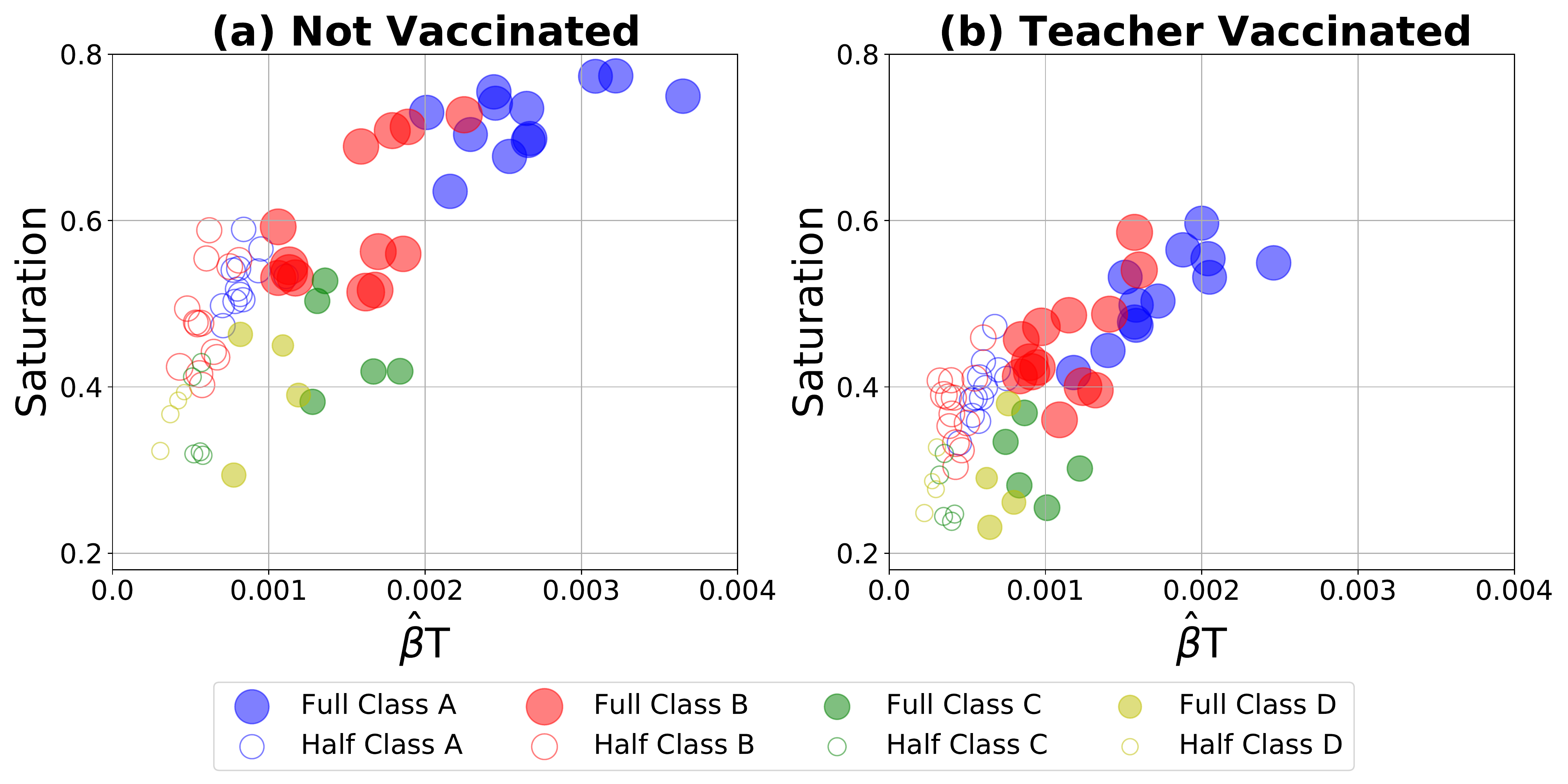}
\caption{Scatter plot between the transmission likelihood $\hat{\beta}T$ and saturation, where each of the 68 circles represents a full-class or a half-class simulation of one of 34 observations in four classes for not vaccinated and teacher vaccinated scenarios \Revised{(See Table 1)}. The area of each circle is proportional to the specific classroom density for a specific observation day, and the color represents Classes A--D for the full and half-sized simulations, respectively.}
\label{fig:saturation}
\end{figure}

Figure~\ref{fig:saturation} plots the transmission likelihood $\hat{\beta}T$ against the saturation for full-class vs. half-class and for not vaccinated vs. teacher vaccinated scenarios. Saturation parameters appeared to grow with transmission likelihood and, at the highest transmission likelihood, converged around $75\%$ and \Revised{$55\%$} for the not vaccinated and teacher vaccinated scenarios, respectively. The area of each circle in Fig.~\ref{fig:saturation} represents the classroom density of the corresponding observation, suggesting a positive correlation between density and saturation. \Revised{The low density classes C and D (Full Class) show similar infection patterns to the half-sized high density classes A and B (Half Class).  It nevertheless may be the case that simulated reductions in classroom density slightly underestimate the effect of actual reduced density, perhaps because simulations do not account for changes in actual social interaction (see SM Section 4.F).}

A mixed effects regression model, with observations nested in classes, predicted saturation and transmission likelihood from class size and vaccination (see Fig.~\ref{fig:saturation}). Half-sized scenarios were associated with a \Revised{18.2\% (Range: 16.0\% - 20.3\%) and 62.9\% (Range: 49.72\%. - 70.87\%)} mean reduction in saturation and transmission likelihood compared to full-sized scenarios, respectively. These full-size to half-size reductions in transmission likelihood, $B=-.001, se=.00006, t=-17.73, p<.00001$, and in mean saturation levels, \Revised{$B=-.11, se=.01, t=-10.21, p<.00001$}, were statistically significant. Likewise, the teacher-vaccinated scenario was associated with a \Revised{25.3\% (Range: 23.4\% - 27.2\%) and 29.4\% (Range: 6.72\% - 45.37\%)} mean reduction in saturation and transmission likelihood compared to the not vaccinated scenario. Effects of vaccination on lower transmission likelihood, $B=-.0004, se=.00006, t=-6.79, p<.00001$, and mean saturation, \Revised{$B=-.13, se=.01, t=-12.59, p<0.00001$}, were also significant.  \Revised{Of note, the overall 25.3\% reduction for the teacher-vaccinated scenario is a combination of sources: 1) the 13.1\% direct effect of reduced infection of vaccinated teachers,  and 2) the 12.2\% indirect effect of vaccinated teachers infecting fewer individuals (children and other teachers). All statistical analyses were performed at the level of observations ($N=34$), which were nested in classrooms. 
Consequently, the number of simulations performed did not artificially reduce p-values.} See SM Section 3 for additional details.

\Revised{While the simulation uses the same infection parameters for each individual, we observe transmission heterogeneity due to behavioral differences. For example, different patient zeros lead to different infection patterns based on individual differences in contact with others (i.e., variation in $r_1\&\theta_1$) (see SM Section 4.G and Fig. S7 for more details).}

\subsection*{Policy Relevant Outcomes}

\begin{figure}[t]
\centering
\includegraphics[width=0.8\linewidth]{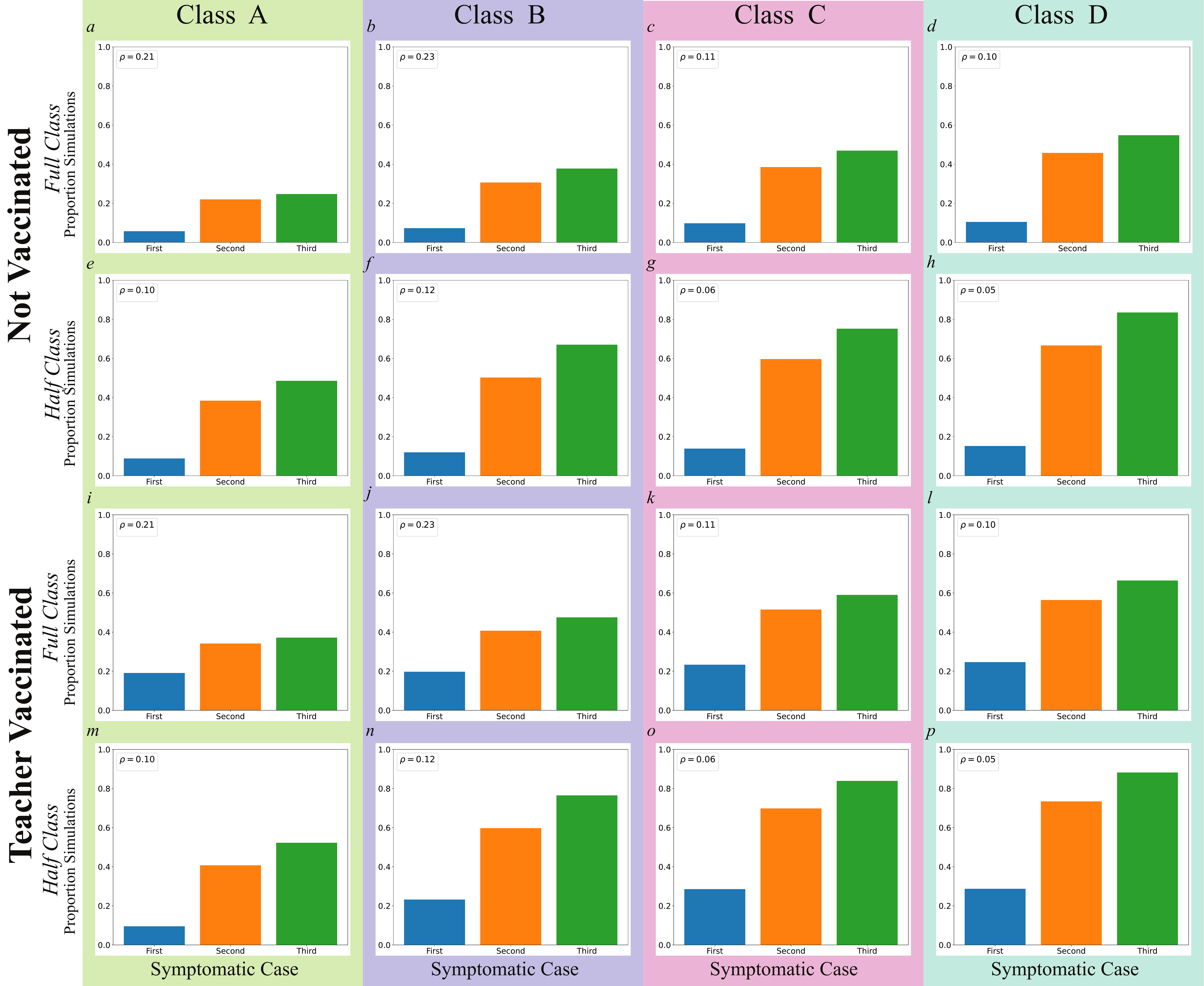}
\caption{The proportion of simulations in which the first, second, and third symptomatic cases do not emerge in classes A-D. Teacher vaccinated and not vaccinated as well as full and half-sized class scenarios are presented. The horizontal axis indicates the nth symptomatic case. The vertical axes indicate the proportion of simulations, and $\rho$'s in each legend represent the classroom densities in the unit of people per $m^2$.}
\label{fig:proportion}
\end{figure}

To explore the criteria under which in-person schooling might be terminated (school shut-down) after discovery of COVID-19 cases, we investigated the timing of the emergence of infected, symptomatic individuals (children or teachers). We first focused on the probability of a given set of simulations yielding a first, second, or third symptomatic individuals (Fig.~\ref{fig:proportion} and Table S3). The $75\%$ symptomaticity rate used in simulations implies that the first symptomatic case is patient zero in three of four simulations. The emergence of the $1^{st}$, $2^{nd}$, and $3^{rd}$ symptomatic case was significantly reduced in the half class scenarios, reflecting sensitivity to classroom density (see Tables S3 and S4). The probability of not detecting a second symptomatic individual, for example, was \Revised{$35.2\%$ and $53.0\%$} for full and half-sized classrooms, respectively. The corresponding probability was $38.7\%$ and \Revised{$49.5\%$} for not vaccinated and teacher vaccinated scenarios, respectively. The reduction associated with teacher vaccination significantly impacted the emergence of a $2^{nd}$, and $3^{rd}$ (but not a $1^{st}$) symptomatic individual (see SM Section 3 and Table S4 for details). 

\begin{figure}[t]
\centering
\includegraphics[width=0.8\linewidth]{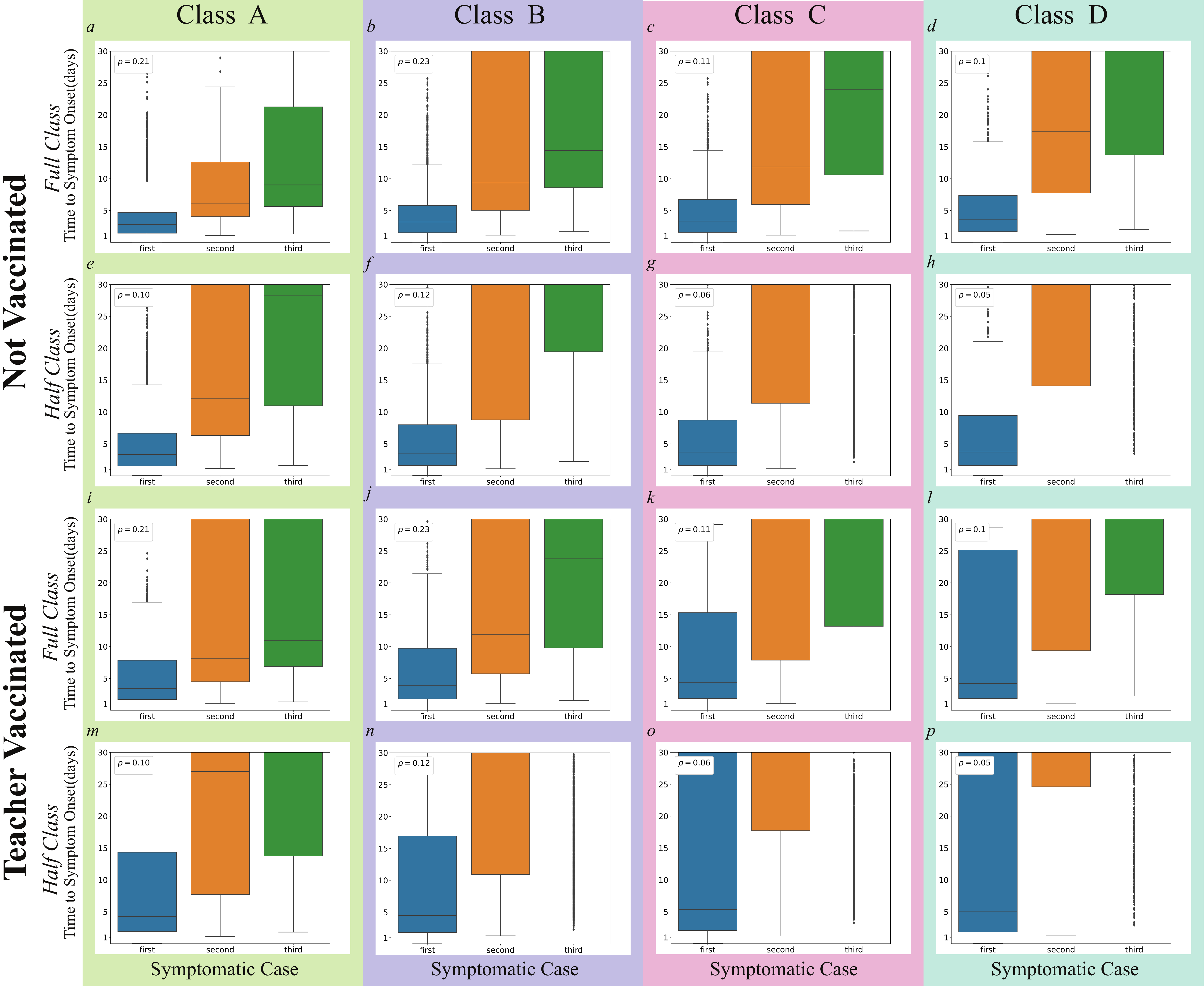}
\caption{Time in days to the emergence of first, second, and third  symptomatic individuals in classes A--D for teacher vaccinated and not vaccinated and for full and half-sized class scenarios. The horizontal axis labels $n^{th}$ symptomatic case. The vertical axis indicates time in days. Horizontal lines within each box indicate medians, and $\rho$'s in each legend represent the classroom densities in the unit of people per $m^2$.}
\label{fig:onset}
\end{figure}

Next we plotted the time in days until the emergence of the first, second, and third symptomatic individual in Fig.~\ref{fig:onset}. The $75\%$ symptomaticity rate again implied that the mean time to the onset of the first symptomatic case typically corresponded to the $4$-day incubation period). Over all \Revised{not vaccinated classrooms}, the median time to the emergence of the second symptomatic individual was \Revised{$8.51$} days in full class and \Revised{$23.00$} days in half class scenarios, reflecting sensitivity to classroom density (Fig.~\ref{fig:onset}). The median time to the emergence of the second symptomatic individual was \Revised{$11.12$} days in the \Revised{teacher vaccinated full-sized class} scenarios and was not observed in \Revised{the teacher vaccinated half-sized class scenarios}. The not-observed value indicates that in over half the teacher vaccinated scenarios a second symptomatic individual never emerged (see Table S5).

\Revised{To explore the impact of classroom behavior, we categorized classroom time as unstructured (free-play and transitions between activities) or as structured. Structured activities were teacher-led and primarily occurred when children were seated at tables (such as circle-time, shared book reading, meal-time, and organized play). Simulations suggest that unstructured time was associated with a higher trajectory of infection than unstructured time, presumably because individuals were in closer proximity and were more mutually oriented during these periods (see SM Section 4.B for more details). We also find that transmission heterogeneity is naturally encoded in our model since individuals behave differently and their social interactions are inhomogeneous (SM Section 4.G).}

\section*{Discussion}

In the face of pandemic threats such as that posed by COVID-19, federal and local governments must adopt policies that weigh the importance of educational opportunities against the risks of infection within classroom settings. School closure reduces transmission risk\cite{honein2021data}, but also limits the employment capacity of health-care and other essential employees who are tasked with child care\cite{o2009if}, 
 which reduces economic productivity\cite{li2020role}. 
Here we explore the impact of less drastic prophylactic measures such as reducing classroom enrollment and teacher vaccination.

Our goal was to describe the development of an agent-based SEIR model to inform decisions and policy on curbing the impact of COVID-19 classroom spread. Existing research examines infection as a product of expected rates of interaction between infected and susceptible individuals, but does not examine the role of individual (agents) in transmission. With few exceptions\cite{stehle2011simulation}, existing models do not describe the actual physical interactions through which individuals infect others. Transmission, however, occurs in physical space over time. 

Since their original formulation\cite{kermack1927contribution}, SEIR models have been used to model infectious transmission at a population level\cite{he2020seir,yang2020modified}. 
The current study’s RFID system provided subsecond ($4 Hz$)  measures of the physical distance and relative orientation of all individuals in a classroom\cite{stehle2011simulation}.  Using the resulting data we employed an agent-based model to examine infection in real-time in a real-life system, the preschool classroom. The model uses actual student and teacher interaction behavior as input and estimates infection probabilities based on interpersonal distance and orientation.

The current policy-focused modeling scenarios indicated that classroom density was a key parameter limiting \Revised{SARS-CoV-2 transmission}. Both simulations that reduced classroom enrollment and simulations of teacher vaccination receipt effectively reduced classroom density by decreasing the number of susceptible individuals (see Fig.~\ref{fig:saturation}). The half class scenario reduced \Revised{overall infection proportion by $18.2\%$} while teacher vaccination was associated with a \Revised{$25.3\%$} reduction. In half class scenarios, the proportion of simulations in which a second symptomatic individual never emerged rose from approximately one third to one half of cases \Revised{($35.2\%$ to $53.0\%$)} while teacher vaccination was associated with a more modest rise \Revised{($38.7\%$ to $49.5\%$)}. The timing of the emergence of these first, second, and third symptomatic cases -- canaries in the coal mine -- was also delayed in both the half class and the teacher vaccination scenarios.

A limitation of the current modeling approach is the need to make assumptions about the physical and temporal parameters characterizing transmission. Modeling parameters assumed that the time lag to infectiousness was one day and the lag to symptomaticity was four days\cite{cdc2019interim}. Likewise, the precise physical parameters involved in \Revised{SARS-CoV-2 transmission} are not known\cite{rosti2020fluid}. The current models, however, allow for flexible modeling of the physical and temporal parameters associated with pathogen spread in enclosed spaces over time. Note, for example, that CDC changes in safe distance recommendation from $6$ to $3$ feet do not change key model parameters\cite{bertozzi2020challenges,van2021effectiveness}. Differential infectiousness of \Revised{SARS-CoV-2 variants}\cite{delta_variant} (e.g., the Delta variant, $B.1.617.2$), changes in vaccine effectiveness and availability, and mask mandates will all affect infection levels, as will community positivity. \Revised{We have explored the impact of parameters to the transmission patterns in SM Section 4.} That key contribution of the models is their flexibility. \Revised{Indeed, as our model has been purposely designed to be modular and therefore used as a general framework, it allows inputs to change to reflect a multitude of future scenarios that include different pathogens such as COVID variants and new viral pathogens, and different transmission vectors including airborne and aerosol particles, as well as variation in class room size/density, and vaccination status.}

\section*{Data Availability}
The datasets generated during and/or analysed during the current study are available in the OSF repository,  \url{https://osf.io/h7ks8/?view_only=5b03a51e79ff4c57b59c2a814d60dbd3}.

\bibliography{reference}

\section*{Acknowledgements}
This work was financially supported by the National Science Foundation, IBSS-L1620294; Institute of Education Sciences, R324A180203; Microsoft AI for Health COVID-19 Grant Program and Google Cloud COVID-19 Research Credits Program.

\section*{Author contributions statement}
C.S. and D.M. defined the overall problem and designed the study.
P.W., D.M., L.P., Y.T. and C.S. conducted the literature review.
D.M., L.P., Y.T., Y.Z. collected and preprocessed the empirical data.
P.W., Y.Z. and C.S. developed the transmission model.
Y.Z., Y.T. and M.S. prepared and conducted numerical simulations.
L.P., Y.Z. and Y.T. performed the statistical analysis and prepared the figures.
P.W., C.S., and D.M. interpreted findings and wrote the manuscript.
All authors provided critical discussion and approved its submission.

\section*{Additional information}

\textbf{Accession codes}: The codes to run the simulations based on the datasets in this study are available at \url{https://anonymous.4open.science/r/Classroom_COVID_Simluation-2E38}.

\noindent The authors declare no competing interests.

\end{document}